# MOLECULAR WEIGHT DISTRIBUTION MODELING IN STEP GROWTH POLYMERIZATION OF OLIGOMERS

**Serhii Kondratov**

Department of Chemical Engineering and Ecology

Volodymyr Dahl East Ukrainian National University

17 John Paul II Street, Kyiv, 01042, Ukraine

*kondratovsa@gmail.com*

**Abstract**

An approach to modeling the oligomer composition distribution function in the irreversible step growth homopolymerization process based on a mixture of oligomers of arbitrary composition is developed. The approach is based on consideration of probabilities of processes in the system, proceeding from Flory's principle and obtaining on this basis an infinite system of differential equations linking mole fractions of each oligomer of a finite mixture with the degree of transformation. It is shown that this system has analytical solutions and the Flory distribution is its special case at the initial geometric distribution. A procedure for the numerical solution of the system of model equations has been developed. It is based on the solution of a finite system followed by extrapolation by a convergent geometric progression to infinity. The developed approach makes it possible to predict the change in the oligomer composition during the process up to a given degree of polymerization

**keywords:** step-growth polymerization, oligomers, distribution function, differential equations, numerical solving,

# Introduction

Reactive oligomers (RO) are polymers with a low degree of polymerization and active terminal functional groups. They are a modern trend in polymer chemistry and technology, as they can serve as versatile building blocks for creating complex polymeric materials and composites. They are also an important class of organic compounds with peculiar and unique [1–4] properties. They are widely used in the synthesis of various polymeric materials, such as epoxy, polyurethane, acrylic, etc., that have many applications in different fields of human activity. The main method of obtaining oligomers is step-growth polymerization (SGP), which involves polycondensation and polyaddition reactions [1–8]. These reactions produce a mixture of products with a broad molecular weight distribution (MWD). The synthesis is usually stopped when the average degree of polymerization reaches a desired value, which depends on the concentration of terminal functional groups.

The molecular weight distribution is the most important factor that affects the properties of polymers and oligomers [1,4,5,9–12]. Controlling the properties of polymers without changing their chemical composition is a challenging problem in polymer chemical physics, materials science, and engineering [12]. Currently, some progress has been achieved in MWD control by using chemical methods that alter the classical mechanism of step polymerization to a mechanism similar to the "living" anionic polymerization [11,13–15], methods of click-chemistry [14,15] and enzymatic catalysis [16]. These methods are suitable for the synthesis of high polymers with a high degree of polymerization and a narrow MWD, but not for oligomers.

Step-growth polymerization processes are usually performed with monomers as the starting materials. However, it is also possible to use low molecular weight oligomers, either in pure form (if such products are accessible) or in mixtures, for the same purpose. Such mixtures of oligomers can be obtained, for example, by extracting polymers with supercritical fluids, especially with carbon dioxide [17–19]. From this perspective, changing the composition of the initial substances in the reactions can be a promising method to control the MWD of oligomers.

Step-growth polymerization processes are usually performed with monomers as the starting materials. However, it is also possible to use low molecular weight oligomers, either in pure form (if such products are accessible) or in mixtures, for the same purpose. Such mixtures of oligomers can be obtained, for example, by extracting polymers with supercritical fluids, especially with carbon dioxide [17–19]. From this perspective, changing the composition of the initial substances in the reactions can be a promising method to control the MWD of oligomers. Mathematical models of step-growth polymerization processes have been extensively studied since the classical works of P. Flory [20, 21]. They have been reviewed and generalized in several publications and monographs [5,21–26]. The foundation for developing models of step polymerization processes is the principle of P. Flory, which states that the reactivity of active centers (terminal functional groups of polymers) does not depend on their position in the chain [20,21]. Two approaches are used for analytical modeling of step-growth polymerization: kinetic and statistical [22–24]. In the first approach, systems of differential equations are derived and solved to describe the kinetics of concentration changes of all types of molecules involved in the process. In the second approach, each macromolecule is considered as a realization of a separate random process of conditional motion along the polymer chain. The probability of this realization is equal to the fraction of molecules corresponding to it among all molecules of the reaction mixture [23]. The "ideal system" is a "test case" for the study of polycondensation processes from the perspective of the theory of irreversible linear homopolycondensation of the form:

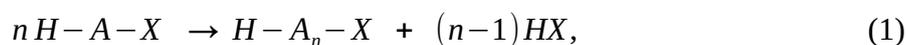

$$n H-A-X \;\to\; H-A_n-X \;+\; (n-1)HX, \qquad (1)$$

where $H,X$ are the fragments of active terminal functional groups that react;
$HX$ are the molecules of low molecular weight products (water, alcohol, halogen hydrogen, etc.), which are released during chain growth reactions. This scheme also corresponds to irreversible polyaddition, where there is no formation of low-molecular-weight products:

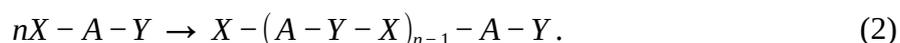

$$nX-A-Y \;\to\; X-(A-Y-X)_{n-1}-A-Y. \qquad (2)$$

The mechanism of the processes is that any two components of the system $H-A_i-X$ and $H-A_j-X$ can collide with each other randomly and, according to Flory's principle, react with the same rate constant $k$:

$$H-A_i-X \ +\ H-A_j-X\ \xrightarrow{k}\ H-A_{i+j}-X\ +\ HX, \quad (\{i,j\}=1,2,\ldots). \tag{3}$$

From this scheme, P. Flory [20] obtained the equation for the distribution of quantities of molecules of polymers with the degree of polymerization $i$ in the step-growth polymerization of monomers:

$$N_i = N_0 \cdot (1-x) \cdot x^{i-1}, \tag{4}$$

where $N_0$ is the initial number of monomer molecules, $N_i$ is the number of polymer molecules with degree of polymerization $i$, $x$ is the degree of monomer conversion in the step-growth polymerization reaction.

The average degree of polymerization $\bar{N}$ for this distribution is described by equation (5):

$$\bar{N} = \frac{\sum_{i=1}^{\infty} i \cdot N_i}{\sum_{j=1}^{\infty} N_j} = \sum_{i=1}^{\infty} i \cdot \pi_i = \frac{1}{1-x}; \quad \pi_i = \frac{i \cdot N_i}{\sum_{j=1}^{\infty} N_j}, \tag{5}$$

where $\pi_i$ is the mole fraction of oligomer with degree of polymerization $i$ in the mixture.

In the kinetic approach, an infinite-dimensional system of ordinary differential equations of kinetics is generated for each component [22-24]:

$$\frac{dx_n}{d\tau} = \frac{1}{2} \cdot k \cdot \sum_{j=1}^{n-1} x_j \cdot x_{n-j} - k \cdot x_n \cdot \sum_{j=1}^{\infty} x_j \quad (n=1,2,\ldots) \tag{6}$$

In equation (6), the first sum represents the rate of accumulation of an oligomer with a degree of polymerization $n$ along each possible route, and the second sum represents the rate of consumption of this oligomer along all possible routes. To simplify and analytically solve the system of equations (6), the method of generating functions is used [22-29]. In this method, the real-time $t$ is replaced by the reduced time $\tau$:

$$t = \int_0^{\tau} k \cdot d\tau, \tag{7}$$

This allows us to eliminate the rate constant in the system (6). The discrete distribution of oligomer concentrations $C(l)$ is replaced by a continuous generating function of the form:

$$g(\xi) = \sum_{l=1}^{\infty} C(l) \cdot \xi^l. \tag{8}$$

The argument of the generating function $\xi$ is purely auxiliary and does not correspond to any physical quantity. If the distribution $C(l)$ also depends on other parameters, then the function $g(\xi)$ will also depend on them. The generating function uniquely determines the distribution of $C(j)$: the values of $C(j)$ are the coefficients of the Taylor series expansion of $g(\xi)$ around $\xi = 0$ and can be obtained as derivatives at this point [22,23]:

$$C(l) = \frac{1}{j!} \cdot \frac{d^j g(\xi)}{d \xi^j} \bigg|_{\xi=0} \; ; \quad j=1,2,\ldots \quad (9)$$

When solving, the system of infinite-dimensional ordinary differential equations of kinetics is replaced by one first-order partial differential equation with respect to the generating function. Its integration gives a generating function that depends on time and $\xi$ [5, 23, 30]. Next, the distribution function can be derived from the generating function.

The method of generating functions is currently the main theoretical method for investigating the distribution functions of linear and branched polymers based on monomers [5,22–30]. However, in this case, the problems often cannot be solved analytically and one has to resort to numerical solutions [31–33]. In the case of step growth polymerization of oligomers, another problem arises. For each initial composition of the oligomer mixture, we have to solve a nontrivial problem, i.e., to construct own generating function. Therefore, a different approach is needed to model the step-growth polymerization of oligomers. The present work is concerned with the development of such an approach.

## 2. Model

Let us consider the process of irreversible ideal homopolycondensation, in which the individual reactions proceed according to scheme (3). We assume that the system follows the Flory principle of independence of reactivity of polymer terminal groups from chain length and that there are no side reactions that prevent the formation of polymers. As is known [5], the rate constant of polycondensation is proportional to the probability of group collisions. Therefore, we consider that the probability of interaction between the end groups $H$ and $X$ of oligomers $H - A_i - X$ and $H - A_j - X$ ($\{i,j\} = 1,2,\ldots$) does not depend on the length of the polymer chain. Let's suppose that at the initial moment in time, the system has $N_i^0$ molecules of oligomer with a degree of polymerization ($DP$) $i = 1, 2, \ldots M_0$, ($M_0$ is the highest degree of polymerization of the oligomer in the initial mixture). In each reaction (3), the number of molecules decreases by one, so after $m$ such reactions, the number of molecules of all oligomers will be:

$$N_0^{(m)} = N_0 - m. \quad (10)$$

Let us consider the system:
- $M_m$ is the highest degree of polymerization of the oligomer in the mixture after $m$ process steps;
- $N_0^i, N_i$ are the number of molecules with $DP = i$ at the initial time and after $m$ process steps;
- $N_0, N_0^m$ are the number of molecules of all polymers (oligomers) in the system at the initial moment of time and after $m$ steps of the process:

$$N^0 = \sum_{i=1}^{M_0} N_i^0, \quad N_0^{(m)} = \sum_{i=1}^{M_m} N_i; \quad (11)$$

- $\pi_i$ is the fraction of oligomer molecules with the degree of polymerization $i$ concerning the total number of molecules at the initial time:

$$\pi_i = \frac{N_i}{N^0}; \qquad (12)$$

- the system is balanced by the number of repeating elementary units:

$$\sum_{i=1}^{M_m} i \cdot N_i = \sum_{i=1}^{M_0} i \cdot N_i^0; \qquad (13)$$

- the number-average degree of polymerization after $m$ steps is:

$$\bar{N}_m = \frac{\sum_{i=1}^{M_m} i \cdot N_i^0}{N_i^0}. \qquad (14)$$

At the $(m+1)$-th step the following events related to this degree of polymerization are possible:
- <u>event A1</u>: the number of oligomer molecules with $DP=i$ will increase by 1. The probability of the event is $p_{+1}^m$;
- <u>event A2</u>: the number of oligomer molecules with $DP=i$ will decrease by 1. The probability of the event is $p_{-1}^m$;
- <u>event A3</u>: the number of oligomer molecules with $DP=i$ will decrease by 2. The probability of the event is $p_{-2}^m$;
- <u>event A4</u>: the number of oligomer molecules with $DP=i$ will not change. The probability of of the event )is $p_0^m$.

These events form a complete group, the sum of their probabilities is 1. Hence:

$$p_0^{(m)} = 1 - p_{+1}^{(m)} - p_{-1}^{(m)} - p_{-2}^{(m)}. \qquad (15)$$

The average number of molecules with degree of polymerization $i$ at step number $m+1$ with regard to (15) is:

$$\begin{aligned} N_i^{m+1} &= \left(N_i^m+1\right) \cdot p_{+1}^{(m)} + \left(N_i^m-1\right) \cdot p_{-1}^{(m)} + \left(N_i^m-2\right) \cdot p_{-2}^{(m)} + N_i^m \cdot p_0^{(m)} = \\ &= \left(N_i^m+1\right) \cdot p_{+1}^{(m)} + \left(N_i^m-1\right) \cdot p_{-1}^{(m)} + \left(N_i^m-2\right) \cdot p_{-2} + N_i^m \cdot \left(1 - p_{+1}^{(m)} - p_{-1}^{(m)} - p_{-2}^{(m)}\right) = \\ &= \left(N_i^m+1\right) \cdot p_{+1}^{(m)} + \left(N_i^m-1\right) \cdot p_{-1}^{(m)} + \left(N_i^m-2\right) \cdot p_{-2}^{(m)} + N_i^m \cdot \left(1 - p_{+1}^{(m)} - p_{-1}^{(m)} - p_{-2}^{(m)}\right) = \\ &= N_i^m + p_{+1}^{(m)} - p_{-1}^{(m)} - 2 \cdot p_{-2}^{(m)} \end{aligned} \qquad (16)$$

The increment in the number of molecules with degree of polymerization $i$ will be:

$$\Delta N_i = N_i^m + 1 - N_i^m = p_{+1}^{(m)} - p_{-1}^{(m)} - 2 \cdot p_{-2}^{(m)} \qquad (17)$$

Let us divide both sides of (17) by the smallest possible change in the number of molecules $\Delta m = 1$. The total number of molecules $N_0$ in a polymer system is always very large. Therefore, we can assume that $N_0 \to \infty$. Then the ratio of changes can be regarded as the derivative of $m$:

$$\lim_{N_0 \to \infty} \left[ \frac{\Delta N_i}{\Delta m = 1} = \frac{\dfrac{\Delta N_i}{N_0}}{\dfrac{m}{N_0}} = \frac{d(N_i/N_0)}{d(m/N_0)} \right] = \frac{dN_i}{dm} = \frac{p_{+1}^{(m)} - p_{-1}^{(m)} - 2 \cdot p_{-2}^{(m)}}{(\Delta m = 1)} \qquad (18)$$

As a result, we obtain an infinite system of ordinary differential equations:

$$\frac{dN_i}{dm} = p_{+1}^{(m)} - p_{-1}^{(m)} - 2 \cdot p_{-2}^{(m)}, \quad i = 1, 2, \ldots . \qquad (19)$$

Let us express the probabilities in (19) through the number of molecules with $DP = i$ at step $m$. The event A1 is obtained by the union of mutually exclusive events, each of which represents the intersection of two independent events - the random selection of two oligomers: the first with $DP = j$, and the second with $DP = i - j$; ($j = 1, 2, \ldots, i-1$). According to the rules of combinatorics and probability theory, when $N_0 - m \gg 1$:

$$p_{+1}^{(m)} = \sum_{j=1}^{i-1} \left[ \frac{N_j}{(N_0 - m)} \cdot \frac{N_{i-j}}{(N_0 - m - 1)} \right] \approx \sum_{j=1}^{i-1} \frac{N_j \cdot N_{i-j}}{(N_0 - m)^2} \qquad (20)$$

The event A2 consists of the union of incompatible events: the random choice of the first oligomer with $DP = i$, the second oligomer with $DP \neq i$, or the first oligomer with $DP \neq i$, the second oligomer with $DP = i$. Since $\sum_{j=1}^{m} N_j = N_0 - m$, when $(N - m) \gg 1$:

$$p_{-1}^{(m)} = \frac{2 \cdot \left( \sum_{j=1}^{i-1} N_i \cdot N_j + \sum_{j=i+1}^{M_m} N_i \cdot N_j \right)}{(N_0 - m) \cdot (N_0 - m - 1)} \approx \frac{2 \cdot \sum_{j=1}^{M_m} N_i \cdot N_j}{(N_0 - m)^2} - \frac{2 \cdot N_i^2}{(N_0 - m)^2} = \\ = \frac{2 \cdot N_i \cdot \sum_{j=1}^{M_m} N_j}{(N_0 - m)^2} - \frac{2 \cdot N_i^2}{(N_0 - m)^2} . \qquad (21)$$

From (10) and (11):

$$\sum_{j=1}^{M_m} N_j = N_0^m = N^0 - m \quad , \qquad (22)$$

get:

$$p_{-1}^{(m)} = \frac{2 \cdot N_i \cdot \sum_{j=1}^{M_m} N_j}{(N_0 - m)^2} - \frac{2 \cdot N_i^2}{(N_0 - m)^2} = \frac{2 \cdot N_i}{(N_0 - m)} - \frac{2 \cdot N_i^2}{(N_0 - m)^2} . \qquad (23)$$

Event A3 consists of the union of incompatible events: random selection of the first and second monomer with $DP = i$. The probability of this event at $N_0 - m \gg 1$ is:

$$p_{-2}^{(m)} = \frac{N_i^2}{(N_0-m)\cdot(N_0-m-1)} \approx \frac{N_i^2}{(N_0-m)^2} \qquad (24)$$

Let us substitute (20, 23, 24) into (19) and give similar ones. We obtain the differential equation of the dependence of the number of monomer molecules with $DP=i$ as a function of the number of process steps:

$$\frac{dN_i}{dm} = \frac{\sum_{j=1}^{i-1} N_j \cdot N_{i-j}}{(N_1^0-m)^2} - \frac{2\cdot N_i}{(N_1^0-m)}. \qquad (25)$$

Let's do the substitutions:

$$x=\frac{m}{N_0}, \quad \pi_i=\frac{N_i}{N_0-m}; \qquad (26)$$

$x$ is the ratio of the number of steps of the stepwise polymerization process to the total number of molecules $N_0$, $\pi_i$ - the fraction of oligomer molecules with $DP=i$ in the mixture after $m$ steps. According to (21), $N_0$ is the limiting number of steps that must be performed for complete polymerization and formation of a single macromolecule. Therefore, $x$ can be regarded as the fraction of polymerization steps relative to the limiting number of steps. In notation (26):

$$N_i = \pi_i \cdot N_0 \cdot (1-\frac{m}{N_0}) = \pi_i \cdot N_0 \cdot (1-x), \quad \frac{dx}{dm} = \frac{1}{N_0}, \qquad (27)$$

$$\frac{dN_i}{dx} = N_0 \cdot \left[(1-x)\cdot\frac{d\pi_i}{dx} - \pi_i\right], \qquad (28)$$

$$\frac{dN_i}{dm} = \frac{dN_i}{dx}\cdot\frac{dx}{dm} = (1-x)\cdot\frac{d\pi_i}{dx} - \pi_i. \qquad (29)$$

As a result, when $M_m \to \infty$, we obtain an infinite system of ordinary linear differential equations of the first order:

$$\frac{d\pi_i}{dx} = \frac{\sum_{j=1}^{i-1}\pi_{(i-j)}\cdot\pi_j - \pi_i}{1-x}, \quad (i=1,2,\ldots,M_m) \qquad (i=1,2,\ldots,M_m) \qquad (30)$$

The initial conditions for this system are the composition of the oligomer mixture at $x=0$:

$$\pi_1(0)=\pi_1^0, \quad \pi_2(0)=\pi_2^0, \quad \ldots, \quad \pi_{M_m(0)}=\pi_{M_m}^0, \qquad (31)$$

where $\pi_i^0 = N_i^0/N_0$ is the mole fraction of oligomers with $DP=i$ in the mixture before the start of the process. If there is no oligomer with $DP=i$ in the system at the initial moment, then $\pi_i^0=0$. By definition:

$$\sum_{i=1}^{M_m} \pi_i^0 = 1; \quad 1 \geq \pi_i^0 \geq 0 \, (i=1,2,\ldots,M_m,\ldots). \tag{32}$$

## 3. Exploring the model

The system (30) is recurrent: the unknowns are determined by successively solving 1,2,…,n… equations. The arbitrary $i$-th equation of the system (30) contains $\pi$ values only with numbers from 1 to $i$ and does not contain unknowns with higher numbers. Therefore, each partial solution contains initial conditions only for unknowns with numbers not exceeding $i$. As a consequence, when the dimensionality of the system (30) is truncated to a finite value $n$ for any $i \leq n$, the values of $\pi_i$ from 1 to $n$ and their sum will not change with further increase of the dimensionality of the problem.

Let us prove that any equation of the system (30) is integrable in quadrature and can be solved analytically. Let us consider successively the first three equations of this system and their solutions. The first equation of the system (30) has the form:

$$\frac{d\pi_1}{dx} = -\frac{\pi_1}{1-x}, \quad \pi_1(0) = \pi_1^0, \tag{33}$$

This is an equation with separating variables, the solution is of the form:

$$\pi_1 = \pi_1^0 \cdot (1-x). \tag{34}$$

The second equation of the system (30) has the form:

$$\frac{d\pi_2}{dx} = \frac{\pi_1^2 - \pi_2}{1-x} = \frac{[\pi_1^0]^2 \cdot (1-x)^2 - \pi_2}{1-x}, \quad \pi_2(0) = \pi_2^0. \tag{35}$$

$$\frac{d\pi_2}{dx} + \frac{\pi_2}{1-x} = [\pi_1^0]^2 \cdot (1-x). \tag{36}$$

Equation (36) is a first order linear differential equation that can be solved analytically by the integrating factor method [34]. The integrating factor and the solution for (36) are of the form:

$$\mu = \exp\left(\int \frac{dx}{1-x}\right) \tag{37}$$

$$\pi_2 = \frac{1}{\mu} \cdot \left[\int \mu \cdot (\pi_1^0)^2 \cdot (1-x) dx + \pi_2^0\right] = \left[(\pi_1^0)^2 \cdot x + \pi_2^0\right] \cdot (1-x) \tag{38}$$

After substituting (34) and (38), the third equation of the system (30) will take the form of:

$$\frac{d\pi_3}{dx} = \frac{2 \cdot \pi_1 \cdot \pi_2 - \pi_3}{1-x} = \frac{2 \cdot Q_0(x) \cdot Q_1(x) \cdot (1-x)^2 - \pi_3}{1-x} = -\frac{\pi_3}{1-x} + \\ + 2 \cdot Q_0(x) \cdot Q_1(x) \cdot (1-x) = -\frac{\pi_3}{1-x} + 2 \cdot \pi_1^0 \cdot \left[(\pi_1^0)^2 \cdot x + \pi_1^0\right] \cdot (1-x) \tag{39}$$

This equation has a structure similar to (36). In this equation, the integrating factor is given by (37), and the term on the right-hand side of the equality sign is a quadratic polynomial that has a factor of $1-x$, which is the reciprocal of the integrating factor (37). Therefore, the solution of the third equation of the system is the product of the quadratic polynomial and $1-x$. By extending the same reasoning to the subsequent equations of the system (30), we can also show that any arbitrary equation with number *i* is a product:

$$\pi_i = \left[ \int \left( \sum_{j=0}^{i-2} Q_j(x) \cdot Q_{i-j}(x) \right) \cdot dx + \pi_i^0 \right] \cdot (1-x) \tag{40}$$

The product of polynomials in (40) is also a polynomial, which can be integrated by using the method of quadrature. Therefore, any of the solutions of the system can be expressed as an analytic expression, a polynomial whose coefficients are functions of the initial conditions.

We will prove that an infinite sequence of positive solutions $\pi_i$ of the system (30) forms a convergent numerical series:

$$y = \sum_{i=1}^{\infty} \pi_i = 1 \tag{41}$$

Let us summarize all equations of the system (30):

$$\frac{dy}{dx} = \sum_{i=1}^{\infty} \frac{d\pi_i}{dx} = \frac{1}{1-x} \cdot \left[ \sum_{i=1}^{\infty} \sum_{j=1}^{i-1} \pi_i \cdot \pi_{i-j} - \sum_{i=1}^{\infty} \pi_i \right] = \frac{1}{1-x} \cdot \sum_{i=1}^{\infty} \left[ \pi_i^2 - 2 \cdot \pi_i \cdot \sum_{j=1}^{i-1} \pi_j - \pi_i \right] \tag{42}$$

Since

$$\sum_{i=1}^{\infty} \pi_i^2 + \sum_{i=1}^{\infty} \sum_{i=1}^{i-1} 2 \cdot \pi_i \cdot \pi_j = \left( \sum_{i=1}^{\infty} \pi_i \right)^2 = y^2, \tag{43}$$

equation (42) is transformed into:

$$\frac{dy}{dx} = \frac{y^2 - y}{1-x}. \tag{44}$$

By definition, the initial condition for this equation is: $y(0) = \sum_{i=1}^{\infty} \pi_i^0 = 1$. For this condition, the implicit partial solution of (44) is:

$$\frac{(y-1) \cdot (1-x)}{y} = 0. \tag{45}$$

The fulfillment of equality (45) for any values of $x \in [0;1]$ is possible only if condition (41) is satisfied.

It is interesting to consider the case when, at an initial time instant, the system has a set of oligomers distributed in the Flory [20, 21]:

$$\pi_i^0 = (1-a) \cdot a^i, \quad i = 1, 2, \dots. \tag{46}$$

The denominator $a \in [0; 1]$ of the geometric progression (46) is the degree of transformation of the monomer into the oligomer mixture, $1-a$ is the fraction of the monomer in the oligomer mixture. The special case $a=0$ means that only a monomer is present in the mixture. Let us analyze the first 4 solutions of the system (30) (equations (34), (38), (47-48)) under initial conditions (46) and the ratio of solutions $\pi_i / \pi_{i-1}$ (Table 1).

$$\pi_3 = (1-x) \cdot [(\pi_1^0)^3 \cdot x^2 + 2 \cdot \pi_1^0 \cdot \pi_2^0 \cdot x + \pi_3^0], \tag{47}$$

$$\pi_4 = (1-x) \cdot [(\pi_1^0)^4 \cdot x^3 + 3 \cdot (\pi_1^0)^2 \cdot \pi_2^0 \cdot x^2 + (2 \cdot \pi_1^0 \cdot \pi_3^0 + [\pi_2^0]^2) \cdot x + \pi_4^0]. \tag{48}$$

Table 1: Equations for mole fractions of π$i$ oligomers with $i = 1 \div 4$ and relations $\pi_i / \pi_{i-1}$

| № | $\pi_i$ | $\pi_i / \pi_{i-1}$ |
|---|---|---|
| 1 | $(1-a)(1-x)$ | - |
| 2 | $(1-x) \cdot (1-a) \cdot [(1-a) \cdot x + a]$ | $(1-a) \cdot x + a$ |
| 3 | $(1-x) \cdot (1-a) \cdot [(1-a) \cdot x + a]^2$ | $(1-a) \cdot x + a$ |
| 4 | $(1-x) \cdot (1-a) \cdot [(1-a) \cdot x + a]^3$ | $(1-a) \cdot x + a$ |

From Table 1, for $i = 2, 3, 4$ the ratio $\pi_i / \pi_{i-1}$ is constant. This is characteristic of the distribution (4) and suggests that the distribution of mole fractions of oligomers under initial conditions (46) also obeys the geometric Flory distribution with the degree of transformation (denominator of the progression) $(1-a) \cdot x + a$:

$$\pi_i = (1-a) \cdot (1-x) \cdot [(1-a) \dot x + a]^{i=1} \tag{49}$$

Let us prove this statement by the method of mathematical induction. Let the distribution (49) be satisfied for numbers 1 to $i-1$. Then the differential equation of the system (30) for the number $i$ has the form:

$$\frac{d\pi_i}{dx} = \frac{\sum_{j=1}^{i-1} \pi_{(i-j)} \cdot \pi_j - \pi_i}{1-x} = \frac{(1-a)^2 \cdot (1-x)^2 \cdot \sum_{j=1}^{i-1} [(1-a) \cdot x + a]^{i-j-1} \cdot [(1-a) \cdot x + a]^{j-1} - \pi_i}{1-x} = \tag{50}$$

$$= \frac{(i-1) \cdot (1-a)^2 \cdot (1-x)^2 \cdot [(1-a) \cdot x + a]^{i-2} - \pi_i}{1-x}.$$

Its solution:
$$\pi_i = (1-x) \cdot \{(1-a) \cdot [(1-a) \cdot x + a]^{i-1} + C\} \tag{51}$$

To calculate the integration constant $C$ we use the initial conditions (46):

$$\pi_i|_{x=0} = (1-a) \cdot a^{i-1} + C = (1-a) \cdot a^{i-1}; \quad C = 0 \tag{52}$$

Under this condition, equation (51) coincides with (49). This proves the existence of a geometric distribution for the initial distribution (46) for any number $i$. Earlier [29] a similar result was also obtained using the method of generating functions.

The results obtained show that the distribution described by the system (30) is general for step growth polymerization, and the Flory distribution is its special case when the process is carried out with pure monomers.

## 4. Numerical modeling

In the general case, there is no recurrence formula for $\pi_i$ that can be derived for the solution. As the degree of polymerization $i$ increases, the formulas for calculating the values of $\pi_i$ under arbitrary initial conditions that satisfy (32) become very complicated. Therefore, numerical methods should be used to calculate these values in general. According to the problem statement, to find the distribution function $\pi_i$ $(i=1,2,\ldots)$, it is necessary to find the value of $x \in (0;1)$ that makes the average degree of polymerization differ from the given $\bar{N}_{Giv}$ by less than a given small $\varepsilon > 0$:

$$\left| \sum_{i=1}^{\infty} i \cdot \pi_i(x) - \bar{N}_{Giv} \right| < \varepsilon \tag{53}$$

The main challenge is to solve the system (30) for a given value of $x$ and to compute the infinite sum in (53) based on it. The algorithm is based on the following considerations. The problem is related to the determination of the distribution function of oligomers with bounded $DP$ under condition (32). As the degree of polymerization $i$ increases, the value of $\pi_i \to 0$. It can be assumed that starting from a certain number $n$, for $i > n$, the change in the oligomer fraction will follow the law of a convergent geometric progression:

$$\pi_i = A \cdot b^{i-n-1}, \ A>0, \ 0<b<1. \tag{54}$$

After logarithmization, we obtained a linear equation:

$$\log(\pi_i) = \log(A) + (i-n-1) \cdot \log(b) \tag{55}$$

The parameters $\log(A)$ and $\log(b)$ can be calculated from the last 10-15 values of $\pi_i$ from equation (55) by the least squares method. Numerical experiments showed that for sufficiently large $n$, the linear dependence (55) is fulfilled with a coefficient of determination $R^2 > 0.999$. On this basis, we calculated the number-average degree of polymerization $\bar{N}$ with extrapolation from (55):

$$\bar{N} = \sum_{i=1}^{\infty} i \cdot \pi_i \approx \sum_{i=1}^{n} i \cdot \pi_i + \sum_{i=n+1}^{\infty} i \cdot A \cdot b^{i-1} = \sum_{i=1}^{n} i \cdot \pi_i + \sum_{i=1}^{\infty} i \cdot A \cdot b^{i-1} - \sum_{i=1}^{n} i \cdot A \cdot b^{i-1} \tag{56}$$

A similar extrapolation of the exponential relationship is used in pharmacokinetics to calculate the total area under the time-dependent curve of drug concentration in blood [35].

For geometric progression:

$$\sum_{i=1}^{\infty} i \cdot A \cdot b^{i-1} = A \cdot \frac{d}{db}\left\{\sum_{i=1}^{\infty} b^i\right\} = A \cdot \frac{d}{db}\left\{\sum_{i=0}^{\infty} b^i - 1\right\} = A \cdot \frac{d}{db}\left\{\frac{b}{1-b}\right\} = \frac{A}{(1-b)^2} \tag{57}$$

Hence an approximate expression for the average degree of polymerization:

$$\bar{N} = \sum_{i=1}^{n} i \cdot \pi_i + \left[\frac{A}{1-b} - \sum_{i=1}^{n} i \cdot A \cdot b^{i-1}\right] = \bar{N}_n^0 + \Delta_N; \tag{58}$$

$\bar{N}_n^0 = \sum_{i=1}^{n} i \cdot \pi_i$ - average degree of polymerization calculated by the model (30) at the number of equations $n$;

$\Delta_N$ - extrapolation correction for $n \to \infty$.

Table 2 shows the results of calculations[1] of the average degree of polymerization for the model system, which has a distribution:

$$\pi_i^0 = \begin{cases} 0.1, & i = 1 \div 10 \\ 0, & i > 10 \end{cases}$$

The values of $x = (0.2; 0.4; 0.8; 0.9)$, up to which the system (30) was integrated, were set. The system included $n = 50 \div 1000$ equations, for which the number-average degree of polymerization $\bar{N}_n^0$ was calculated. After that, 10 new values of $\pi_j$ $(j > n)$ were additionally calculated, from which the coefficients $A, b$ of the regression equation (55) and the coefficient of determination $R^2$ were calculated. Further, the correction $\Delta_N$ and the corrected degree of polymerization $\bar{N}$ were calculated using equation (58).

Table 2 shows that in the very unfavorable case with low values of *n* and sufficiently high values of *x*, the results of calculations using model (30) without extrapolation are unsatisfactory. However, for the last 10 values of *DP*, a linear dependence with a coefficient of determination almost equal to 1 is observed. This justifies extrapolation. The use of extrapolation allows for relatively small $n \sim 50 - 200$ and relatively high $x = 0.9$ to obtain $\bar{N}$ values with the same accuracy as at $n = 1000$. The latter can be considered true since the sum of molar fractions of oligomers is practically equal to 1.

From Table 2, the value of $n = 100$ for model (30) followed by extrapolation by equation (57) is sufficient to obtain $\bar{N}$ with accuracy sufficient for practice. This allows us to significantly reduce the duration of calculations. Thus, it was found that for $x = 0.9$ the calculation duration is approximately proportional to $n^3$, and increasing *n* from 100 to 1000 increases the calculation duration by 3 orders of magnitude.

To calculate the distribution function and the value of *x* for a given average degree of polymerization ($\bar{N}_G$) for a given initial oligomer distribution $\pi_{i0}$, the dichotomy method was used to numerically solve equation (59):

$$f(x) = \bar{N}_G - \bar{N}(x) = 0; \quad x \in (0; 1), \ f(0) < 0; \ f(1) > 0 \tag{59}$$

$\bar{N}(x)$ is the mean value calculated by solving the system (30) of $n = 100$ equations and then extrapolating by the method described above.

---

[1] All calculations were performed in the applied mathematics package Scilab (desktop and online versions), using scripts based on the described algorithms.

Table 2: Calculation results of $\bar{N}_n^0$ and $\bar{N}$ as a functions of n and x

| n | $\sum_{i=1}^{n} \pi_i$ | $R^2$ | $\bar{N}_n^0$ | $\Delta_N$ | $\bar{N}$ |
|---|---|---|---|---|---|
| \multicolumn{6}{c}{x=0.9} |
| 50 | 0.599331 | 0.99999 | 13.5538 | 41.4091 | 54.963 |
| 100 | 0.8441801 | 1 | 31.0908 | 23.9082 | 54.999 |
| 200 | 0.976431 | 1 | 49.0271 | 5.97297 | 55.0001 |
| 400 | 0.9994599 | 1 | 54.7563 | 0.24485 | 55.0011 |
| 700 | 0.9999981 | 1 | 55 | 0.00142 | 55.0015 |
| 1000 | 1 | 1 | 55.0017 | 7.1E-06 | 55.0017 |
| \multicolumn{6}{c}{x=0.8} |
| 50 | 0.851136 | 0.99999 | 16.2045 | 11.2877 | 27.4921 |
| 100 | 0.979247 | 1 | 24.8878 | 2.61218 | 27.4999 |
| 200 | 0.9995962 | 1 | 27.409 | 0.09119 | 27.5002 |
| 400 | 0.9999998 | 1 | 27.5003 | 6.6E-05 | 27.5003 |
| 700 | 1 | 1 | 27.5004 | 9.04E-10 | 27.5004 |
| \multicolumn{6}{c}{x=0.4} |
| 50 | 0.999235 | 0.99997 | 9.12293 | 0.04364 | 9.16657 |
| 100 | 0.9999996 | 1 | 9.16663 | 4.6E-05 | 9.16667 |
| 200 | 1 | 1 | 9.16668 | 2.78E-11 | 9.16668 |
| \multicolumn{6}{c}{x=0.2} |
| 50 | 0.9999922 | 0.99977 | 6.87458 | 0.00042 | 6.875 |
| 100 | 1 | 1 | 6.875 | 3.42E-09 | 6.875 |

Calculations were performed until the interval of *x* values containing the solution became less than $10^{-3}$. In this case, the value of the middle of the interval was taken as the root.

## 5. Results and Discussion

Let us consider the meaning of the quantity *x* included in the model (30). As shown above, if only a monomer is present in the initial mixture, the solution of system (30) is the Flory distribution. For it, the value *x* has the sense of the degree of monomer conversion [21]. At arbitrary composition of the initial mixture for polymerization, the value of *x* also characterizes the extent of the polymerization process: at *x* = 0 the initial mixture remains, and at *x*→1 the whole mixture completely transforms into one giant polymer molecule. Therefore, it is reasonable to call this value the conditional degree of conversion. Let us illustrate the possibilities of the proposed method for solving some problems.

**Example 1**. A mixture of a monomer and a dimer is subjected to step-growth polymerization until $\bar{N}=10$ is reached. How does this change the mole fraction distribution function of the oligomers?

Figure 1 shows the simulation results for monomer-dimer ratios of 1:0; 0.4:0.6; 0.2:0.8 and 0:1. At step-growth polymerization of the monomer, the points of dependence of the mole fraction of oligomers on DP lie on a smooth curve described by the Flory distribution. When a dimer is introduced, the dependence acquires a sawtooth character, with the amplitude of the teeth decaying as *DP* increases (Fig. 1). In the limit, when the dimer is polymerized, oligomers with an odd number are absent (curve 4). Thus, when a mixture of oligomers is polymerized, the molecular weight distribution differs significantly from the classical Flory distribution

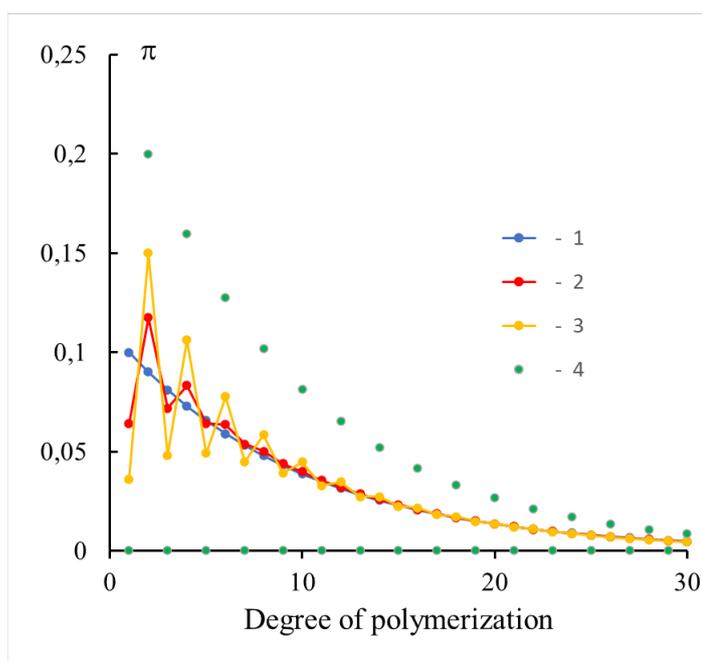

Figure 1: Mole fraction distribution functions of oligomers during stepwise polymerization of monomerdimer blends in mole ratios of 1/0 (1), 0.4/0.6 (2) 0.2/0.8 and 0/1 (3) to an average degree of polymerization of 10.

**Example 2**. A monomer was added to a mixture of oligomers with a Flory distribution and an average degree of polymerization of $\bar{N}=20$. The amount of monomer was *m* moles per mole of the oligomer mixture. The resulting mixture underwent step-growth polymerization until the average degree of polymerization reached $\bar{N}=10$. How did the molecular weight distribution change from the Flory distribution for $\bar{N}=10$?

The simulation results for different oligomer/monomer molar ratios of 0, 0.5, 0.333, and 0.25 (Fig. 2) show that the MWDs obtained from the polymerization also deviated significantly from the Flory distribution for $\bar{N}=10$. Compared to the Flory distribution, the fractions of low molecular weight polymers increased and the fractions of medium molecular weight polymers decreased. Unlike Example 1, in Fig. 2 all data points followed smooth curves.

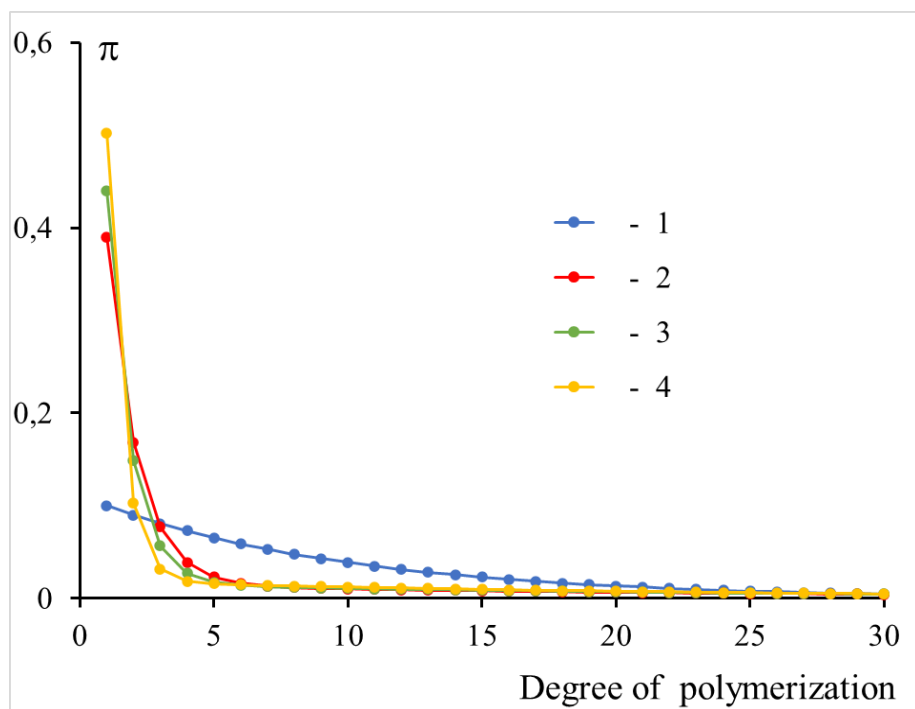

Figure 2: Distribution functions of mole fractions of polymers with $\bar{N}$ = 10 obtained by polymerization of monomer (1) and a mixture of oligomers with $\bar{N}$ = 20 and monomer with mole ratios 1/2 (2), 1/3, (3), 1/4 (4).

**Example 3**. Light oligomers are extracted from a mixture of oligomers with $\bar{N}=5$ distributed over Flory: in the first case 90 % of the monomer, in the second case 90 % of each of the monomer and dimer, and in the third case 90 % of each of the monomer, dimer and trimer. After extraction, the mixtures are further polymerized to $\bar{N}$ = 10. How does this change the distribution from the Flory distribution?

From Fig. 3, when low molecular weight fractions are removed, the distribution of oligomers differs from the Flory distribution: the content of medium molecular weight fractions in the final product increases and the amount of low molecular weight fractions decreases. At the same time, the content of high molecular weight fractions tends to the Flory distribution as the degree of polymerization increases.

# Conclusion

1. A probabilistic approach to modeling the composition of oligomers in the process of irreversible stepwise homopolymerization based on a mixture of oligomers of arbitrary composition has been developed. The approach is based on the consideration of probabilities of processes in the system, proceeding from Flory's principle and obtaining on this basis an infinite system of differential equations relating the mole fractions of each oligomer of a finite mixture to the degree of conversion.

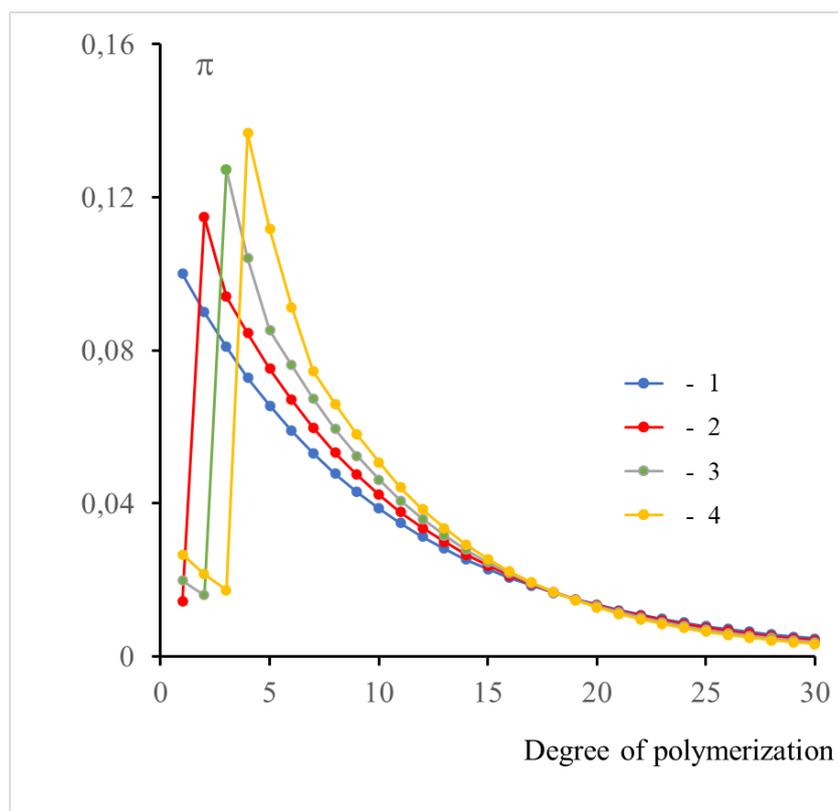

Figure 3: Effect of light fractions extraction on the mole fraction distribution function of oligomers after reaching $\bar{N}=10$. Initial mixture: Flory distribution with $\bar{N}=5$. 1 - without removal of light fractions, 2 - 90 % of monomer removed, 3 - 90 % each of monomer and dimer removed, 4 - 90 % each of monomer, dimer and trimer remove

2. The properties of the model are investigated. It is shown that this system has analytical solutions and the Flory distribution is a special case of the proposed distribution.
3. A procedure for numerical computer solution of the infinite system of equations of the model is developed, based on the solution of the finite system followed by extrapolation by convergent geometric progression to infinity.
4. The developed approach allows predicting possible changes in oligomer composition in a desirable direction by computer simulation.